\documentclass[12pt]{article}
\usepackage[english]{babel}

\usepackage[letterpaper,top=2cm,bottom=2cm,left=3cm,right=3cm,marginparwidth=1.75cm]{geometry}

\usepackage{graphicx,amsmath}
\usepackage{graphicx,amsmath,lineno,color}

\usepackage{units}
\usepackage{caption} 
\usepackage{subcaption}
\usepackage[hidelinks,colorlinks=true,linkcolor=blue,urlcolor=blue,citecolor=blue]{hyperref}
\hypersetup{colorlinks=true}

\newlength{\dinwidth}
\newlength{\dinmargin}
\def\lapproxeq{\lower .7ex\hbox{$\;\stackrel{\textstyle                                                    
<}{\sim}\;$}}                                                    
\def\gapproxeq{\lower .7ex\hbox{$\;\stackrel{\textstyle                                                    
>}{\sim}\;$}}                                                    
\def\be{\begin{equation}}                                                    
\def\ee{\end{equation}}                                                    
\def\bea{\begin{eqnarray}}                                                    
\def\eea{\end{eqnarray}}

\def\sh{\hat s}
\def\sh2{{\hat s}^2}

\begin{document}                                                    
\titlepage                                                    
\begin{flushright}                                                    
%IPPP/19/??  \\                                                    
\today \\                                                    
\end{flushright} 
\vspace*{0.5cm}

\begin{center}                                                    
{\Large \bf The total cross section for proton-proton interactions at the FCC}\\

%Bethe phase variation due to the non-exponential nuclear amplitude $t$ dependence}\\
\vspace*{1cm}
% $\vec\nabla$                                                  

P.~Grafstr\"om  \\                                                   
                                                   
\vspace*{0.5cm}                                                    
 Universit\`a di Bologna, Dipartimento di Fisica , 40126 Bologna, Italy\\
                        
\vspace*{1cm}                                                    
 
\begin{abstract}
The  lower and upper limits of the total cross section  ($\sigma_{tot}$) at the projected FCC-hh have been estimated. A lower limit has been estimated using dispersion relations in combination with recent LHC data of  $\sigma_{tot}$ and the $\rho$-parameter. The upper limit has been estimated using the standard  $ln^{2}(s) $ evolution of $\sigma_{tot}$. Some models giving values in between those limits are also discussed. 

 \end{abstract}

\end{center}

\vspace{1cm}

\section{Introduction}
\label{into}
The Future Circle Collider for hadron collisions i.e. FCC-hh is designed to provide  proton-proton collisions  at a centre-of-mass energy of 100 TeV. At the time   a new accelerator  facility is projected in a previously unexplored energy domain it is  important  to try to  estimate the total interaction probability i.e. the total cross section ($\sigma_{tot}$) at this new energy. A precise calculation is not possible but still one can attempt to estimate an approximate number.  The total cross section  is important for several aspects of an accelerator design like the beam lifetime or heat load on cryogenics, just to mention a few.  In addition radio protection issues and  shielding issues  depend on the  total cross section. Some aspects of detector design might also be of relevance related to the experimental background. Monte-Carlo generators to simulate both signal and background process   often  need as an input the total cross section.

Since the unexpected rise of the  total cross section for pp scattering  was discovered at the ISR it has turned out that at any new accelerator at a higher center-of-mass energy the total cross section has increased as   $ln^{2}(s)$  where $\sqrt{s}$   stands  for the centre-of-mass energy. This was true for the SPS collider, the Tevatron and also  LHC. The Froissart-Martin bound \cite{PhysRev.123.1053}\cite{martin1966extension}  states that  $ \sigma_{tot}$ can not grow faster than $ln^{2}(s)$ with a limit given by  $\sigma_{tot}<\frac{\pi  ^{}}{m^{2}_{\pi^{ ^{}}}}ln^{2}(s)$. However   this is  a bound limiting the maximum possible  asymptotic rise of $\sigma_{tot}$ and not a prediction of an evolution of $\sigma_{tot}$.   Still it seems that up to now $\sigma_{tot}$  actually increases as $ln^{2}(s)$  though with a coefficient  smaller than the  one given by the Froissart-Martin bound. However as will be discussed below in Sect.~\ref{without}, it is not evident any more that a straight forward $ln^{2}(s)$ extrapolation will also work up to 100 TeV.

\begin{figure}
\centering
\includegraphics[width=0.9\textwidth]{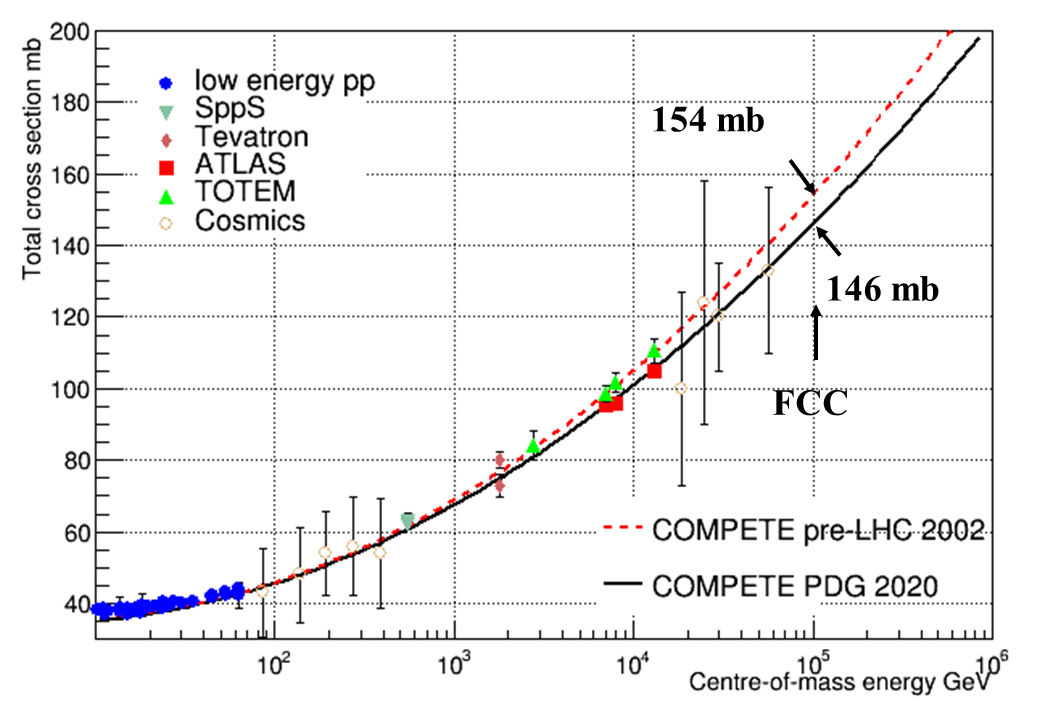}
\caption{\label{fp1}Measurement of the total cross section ($\sigma_{tot}$) as a function of the centre-of-mass energy compared to two $ln^{2}(s)$ parameterizations. More  details are given in the text. }
\end{figure}

Still, assuming that this $ln^{2}(s)$ trend would also continue up to the energy of the FCC,   one would find a $\sigma_{tot}$  around 150 mb.   This is illustrated in Figure~\ref{fp1}  where the world data of $\sigma_{tot}$ is plotted together with two different  $ln^{2}(s)$ predictions. Both have been suggested by the COMPETE collaboration \cite{COMPETE:2002jcr}. The first is taken from their prediction in 2002  i.e. pre-LHC~\cite{COMPETE:2002jcr}.  The  second  $ln^{2}(s)$ parametrisation is taken from the latest  PDG \cite{ParticleDataGroup:2020ssz} which also refers to a  parametrisation proposed by the COMPETE collaboration but in this case in addition using the available LHC data in 2020.
The pre-LHC fit agrees  better with the cross section measured by TOTEM at the LHC while the 2020 fit better match  the ATLAS data. As can be seen there is not a very big difference between the two extrapolations. In one case one obtains 154 mb  and in the other 146 mb at 100 TeV.
  
Block and Halzen~\cite{PhysRevD.72.039902} use an approach very similar to COMPETE. As "ansatz" they use an amplitude giving a    $ln^{2}(s)$   behaviour of   $\sigma_{tot}$ asymptotically and not surprising their result is close to those of COMPETE. At 100 TeV they get 147 mb. 

Probably these kind of estimates  are enough for most practical purposes mentioned above. However this note try to address the question if something more could be said.

\section{Dispersion relations and predictions of the energy evolution of the total cross section}
\label{disp}
A technique to predict the total cross section in a model independent way was pioneered by  the Cern-Rome group at the ISR \cite{Amaldi:1976yf}. It was later also used by the UA4 collaboration at the SppS \cite{UA42:1993ojz}.  The basic idea is to use dispersion relations  to predict  limits of possible    energy evolution  of   $\sigma_{tot}$   above energies where  $\sigma_{tot}$ has been measured. Dispersion relations have been used in physics for more than a century and were introduced in particle physics  by Gell-Mann, Goldhaber and Thirring in 1954~\cite{PhysRev.95.1612}. It is a good example of a possible calculation in the non-perturbative domain based upon very general principles. The corner stones of dispersion relations in particle physics are analyticity of the scattering amplitude, crossing symmetry and unitarity.

By measuring the differential elastic cross section at small angles the   $\rho$-parameter can be extracted. The $\rho$-parameter is defined as the ratio of the real to imaginary part of the elastic scattering amplitude in the forward direction. Dispersion relations connect  the $\rho$-parameter and   $\sigma_{tot}$  in a way that $\rho$ at a given energy becomes sensitive to the energy evolution of $\sigma_{tot}$ beyond that energy. The CERN-Rome group at the ISR showed that in this manner it was possible to predict  $\sigma_{tot}$ at an energy roughly a factor ten higher than the energy at which $\rho$  was measured \cite{Amaldi:1976yf}.

The application of dispersion relations   in this manner is based upon  the general principles mentioned above but  it  is also necessary to assume  that the difference between particle  and antiparticle cross section vanishes asymptotically.

\begin{figure}
\centering
\includegraphics[width=0.9\textwidth]{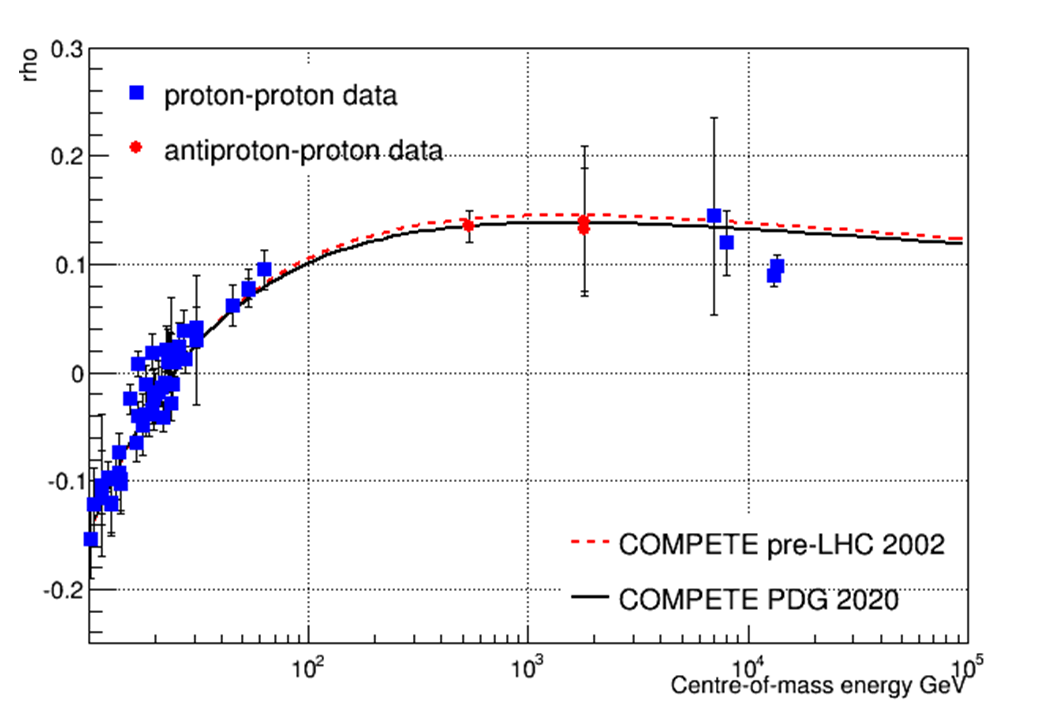}
\caption{\label{fp2} Measurement of the $\rho$-parameter  as a function of the centre-of-mass energy compared to dispersion relations calculations using  two $ln^{2}(s)$ parameterizations of the total cross section ($\sigma_{tot}$) as input. More  details are given in the text. }
\end{figure}

A recent measurement by the  TOTEM  experiment  at the LHC has complicated the situation at LHC energies. The $\rho$-parameter at 13 TeV  was measured and found to be  $\rho=0.09\pm0.01$~\cite{TOTEM:2017sdy}. The ATLAS experiment has later confirmed this measuring~$\rho=0.098\pm0.011$ \cite{arXiv:2207.12246}.  Those values do not agree with predictions  from dispersion relation using the $ln^{2}(s)$ parametrisations of Fig.~\ref{fp1}. This point is illustrated Fig.~\ref{fp2}. The  world data of $\rho$ is plotted together with the prediction of dispersion relations using the COMPETE 2002 parametrisation of  $\sigma_{tot}$ \cite{COMPETE:2002jcr} and also using  the latest COMPETE parameters from the PDG \cite{ParticleDataGroup:2020ssz}.  The deviation between the dispersion relation calculation and $\rho$ is of the order 3-4 $\sigma$  at 13 TeV.

The  most natural explanation of this discrepancy is that  the total cross section at energies beyond LHC grows somewhat slower than $ln^{2}(s).$  Via dispersion relations this would give a lower $\rho$ value. However there is also another option conceivable. 
There is the possibility of the existence  of the so called Odderon. In Regge theory, elastic scattering at high energy is described as an exchange of the  Pomeron in the t-channel. The Pomeron corresponds to a  $CP=++$  state of two gluons in QCD. In the language of QCD the Odderon would be a  $CP=--$ state of three gluons. If the Odderon is present in addition to the Pomeron  the difference between particle and antiparticle total cross sections would not disappear asymptotically and the way to predict $\sigma_{tot}$   using dispersion relation as proposed by the CERN-Rome group would not anymore be possible. 

Thus we have two separate to different scenarios : with and without the Odderon and see in each case what could be said about the energy evolution of $\sigma_{tot}$.

\section{Estimate without the odderon }
\label{without}

As already mentioned, one possible explanation of  a lower $\rho $-value could be that  $\sigma_{tot}$ asymptotically does not follow the  $ln^{2}(s)$   evolution assumed by COMPETE but instead  increases  slightly   slower.  Due to the non-perturbative nature of the problem there is no firm QCD prediction of the energy evolution of $\sigma_{tot}$. A simple way of introducing a slightly slower energy increase  could be to introduce a damping factor of the form \(\frac{ln^{2}(s)}{1+\alpha ln^{2}(s)}\) where \(\alpha\) represents the damping factor.  Such a functional form was proposed in the 80th  by   Block and Cahn~\cite{BLOCK1983224}. They suggested that the damping factor $\alpha $ should be  looked upon as a measure of the energy scale  at which  a possible deviation from  $ln^{2}(s)$ might set in. Bourrely and Martin also used such a damped amplitude in  the context of putting limits on predictions of  $ \sigma_{tot}$ at high energies ~\cite{Bourrely:153114}.
\begin{figure}[ht]
\begin{subfigure}{0.49\textwidth}
\includegraphics[width=0.9\linewidth, height=6cm]{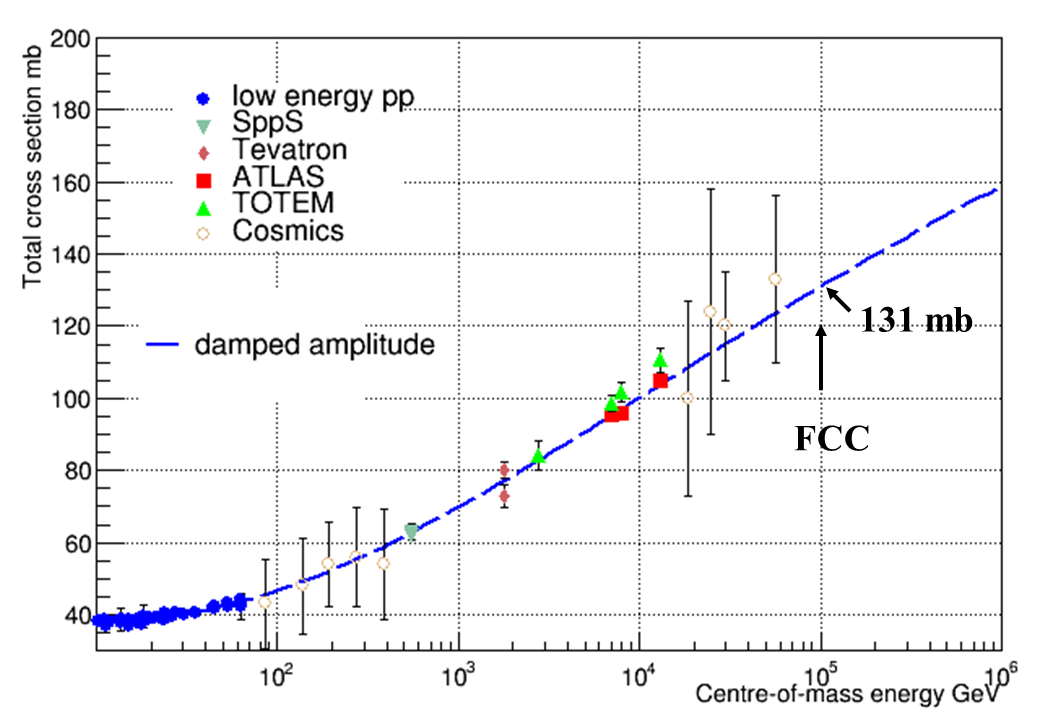} 
\caption{ $\sigma_{tot}$}
\label{fp3}
\end{subfigure}
\begin{subfigure}{0.49\textwidth}
\includegraphics[width=0.9\linewidth, height=6cm]{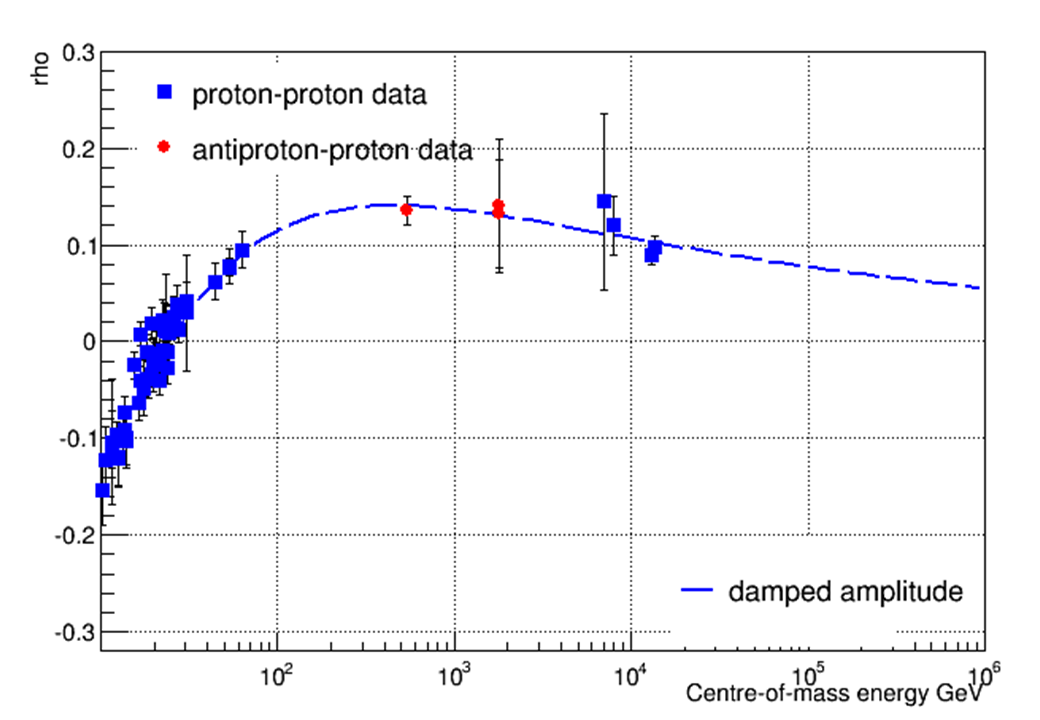}
\caption{$\rho$ }
\label{fp4}
\end{subfigure}
\caption{$\sigma_{tot}$ and $\rho$ as a function of the centre-of-mass energy compared to calculations using a damped scattering amplitude. A damping factor $\alpha=0.0016$ is used. More details are given in the text.}
\label{damped}
\end{figure}
\begin{figure}[ht]
\begin{subfigure}{0.49\textwidth}
\includegraphics[width=0.9\linewidth, height=6cm]{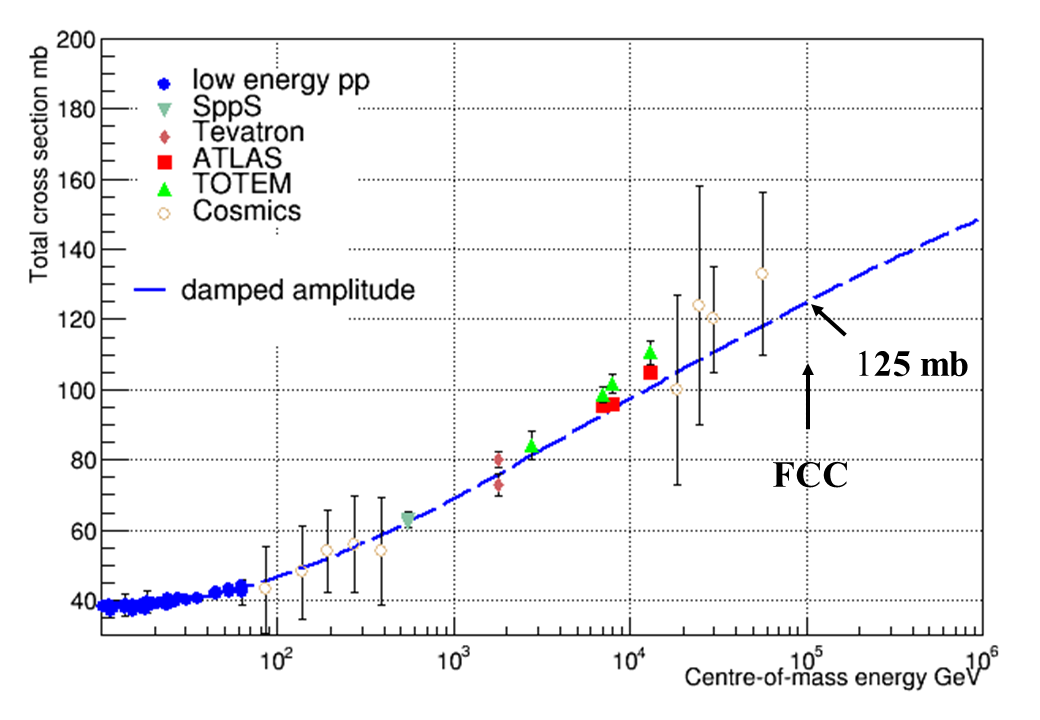} 
\caption{$\sigma_{tot}$}
\label{fp5}
\end{subfigure}
\begin{subfigure}{0.49\textwidth}
\includegraphics[width=0.9\linewidth, height=6cm]{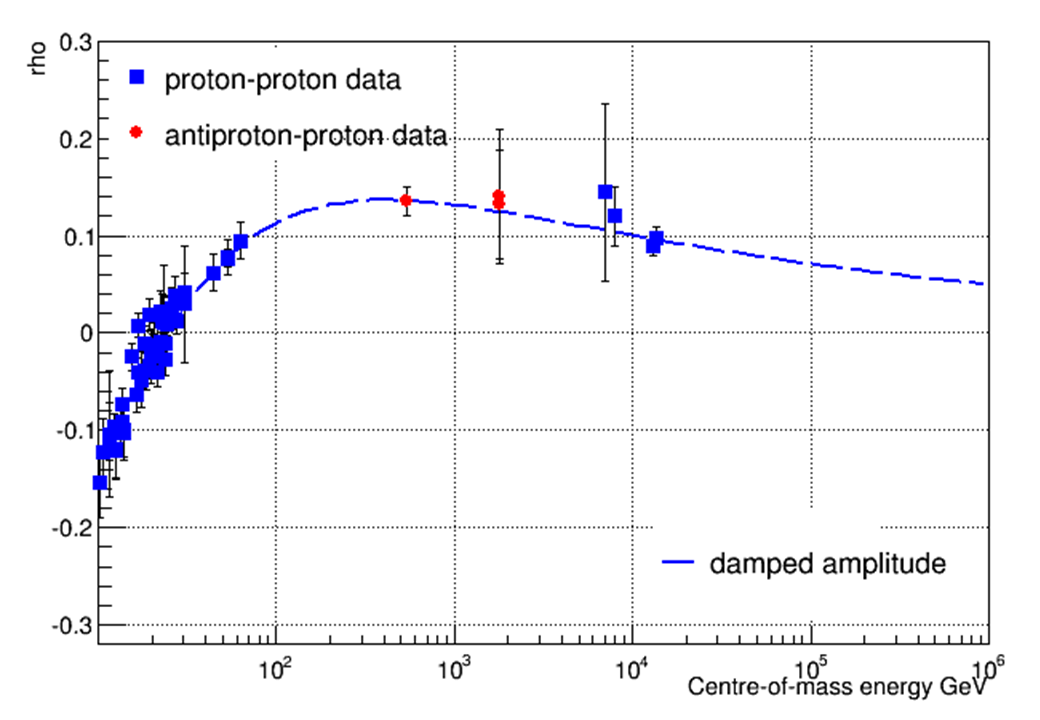}
\caption{$\rho$}
\label{fp6}
\end{subfigure}
\caption{$\sigma_{tot}$ and $\rho$ as a function of the centre-of-mass energy compared to calculations using a damped scattering amplitude. A damping factor $\alpha=0.0019$ is used. More details are given in the text.}
\label{more damped}
\end{figure}

 The proposed amplitude modifies the high-energy behaviour of $\rho$ and  $\sigma_{tot}$. Using this amplitude $ \sigma_{tot}$ will asymptotically approach a constant value. 
We have used dispersion relations and the assumption of a damped amplitude to calculate the corresponding $\rho$-values. The value of the  damping factor has been determined in a global fit of the world data of $\sigma_{tot}$ and $\rho$. The global fit includes all the TOTEM data \cite{TOTEM:2017sdy} at the LHC but not the recent ATLAS data point at 13 TeV  \cite{arXiv:2207.12246}. However including this point would not change the result as the ATLAS data point is completely compatible   with the fit.   The result of the global fit is  shown for $\sigma_{tot}$ and $\rho$  in Fig.\ref{fp3} and  Fig.\ref{fp4}, respectively.   A fair  description of the  data  both for  $\sigma_{tot}$ and $\rho$ is found for a damping factor $\alpha=0.0016$.

From   Fig.\ref{fp3}  can bee seen that $\sigma_{tot}$ at the FCC will be around 130 mb in the case of the damped amplitude and in Fig.\ref{fp4} is seen that such a damped amplitude describes well  the low $\rho$-value found by TOTEM and ATLAS. One might ask the question if the data allow an even stronger damping than what is given by $\alpha=0.0016$. Fig.\ref{fp5} and Fig.\ref{fp6} show the result  for a damping corresponding to  $\alpha=0.0019$. 
The $\rho$-parameter is still well described  but clearly the discrepancy with the measurements of $\sigma_{tot}$ at the LHC starts to be significant. The corresponding  $\sigma_{tot}$ at the FCC is  125 mb.  Thus we conclude from this that 130 mb seems to be a reasonable lower limit of $\sigma_{tot}$ at the FCC.

 It should be stressed that the damped amplitude used here was suggested in the context of putting limits on predictions of  $ \sigma_{tot}$ at high energies and the amplitude was not derived  from an underlying physics model. However there are several attempts to build models describing  the evolution of $\sigma_{tot}$ and $\rho$ based upon more specific physics assumption. 

The Durham group, represented by Khoze, Martin and Ryskin (KMR)
has probably developed the most elaborate approach~\cite{KHOZE2018192}.
They propose a two-channel eikonal model with few parameters and it uses all available  high-energy data of   $\rho$ and  $\sigma_{tot}$ as well as the corresponding  differential elastic cross sections. They also take into account all  available measurements of low mass diffraction. They  use only   a $ C^{+}$ amplitude and thus do not make any  assumption of an additional  Odderon amplitude  and still they get a reasonable description of \(\rho\). This is due to the fact that  $\sigma_{tot}$ increases somewhat slower in resemblance  of the results   using a damped amplitude. Actually the KMR approach gives a good decription of the very recent ATLAS data\cite{arXiv:2207.12246}. This flatter energy dependence is generated by the constraints given by  the data  of  low mass diffraction. At 100 TeV they predict a   $\sigma_{tot}$ of 136 mb.

An alternative to the Regge theory approach is  developed  in Ref.~\cite{PhysRevD.101.074042}. An attempt to describe soft diffraction within the QCD color dipole picture is tried. The dipole amplitude is formulated in an eikonal-like expression. No Odderon signatures are used. The LHC data are well described and the predictions at 100 TeV is 135 mb.

Gotsman and collaborators have  a model based upon on one hand Color Glass Condensate saturation and on the other hand a pure phenomenological treatment of  non-perturbative long distance physics~\cite{GOTSMAN2018155}. At 57 GeV they find a  $\sigma_{tot}$ of 132 mb and a straight forward extrapolation to 100 TeV gives 140 mb.

\section{Estimate with the odderon  }
\label{with}
As mentioned in previous Sect.\ref{disp}~ there is another possibility to generate the  low   $\rho$-value measured  by the TOTEM and ATLAS experiment. A scattering amplitude which is odd under crossing asymptotically i.e. different for particle and antiparticle could also explain the TOTEM/ATLAS result. Such a amplitude would   generate   a lower \(\rho\)-value for pp scattering and a higher value for antiproton scattering relative to  the crossing even amplitude. Those ideas were first formulated in 1973 \cite{Lukaszuk:1973nt}. With the invent of QCD such an amplitude was identified as a three gluon state  $CP=--$  in contrast to the two gluon state $CP=++$ of the Pomeron.

The original idea from 1973 was to  define the  amplitude to be as  strong as allowed by axiomatic field theory. This extreme assumption has as consequence  that not only the imaginary part of the scattering amplitude increases as $ln^{2}(s)$  but also the real  part. This model was later baptized  FMO~\cite{Martynov:2017zjz} (the Froissaron Maximal Odderon). Theoretically there is nothing in QCD that stops the Odderon from existing. However the size of the the contribution is much debated in the literature and the work in Ref. \cite{Donnachie_2022}   even states that  there is lack of evidence for an Odderon contribution at small $t$. 

\begin{figure}[ht]
\begin{subfigure}{0.49\textwidth}
\includegraphics[width=0.9\linewidth, height=6cm]{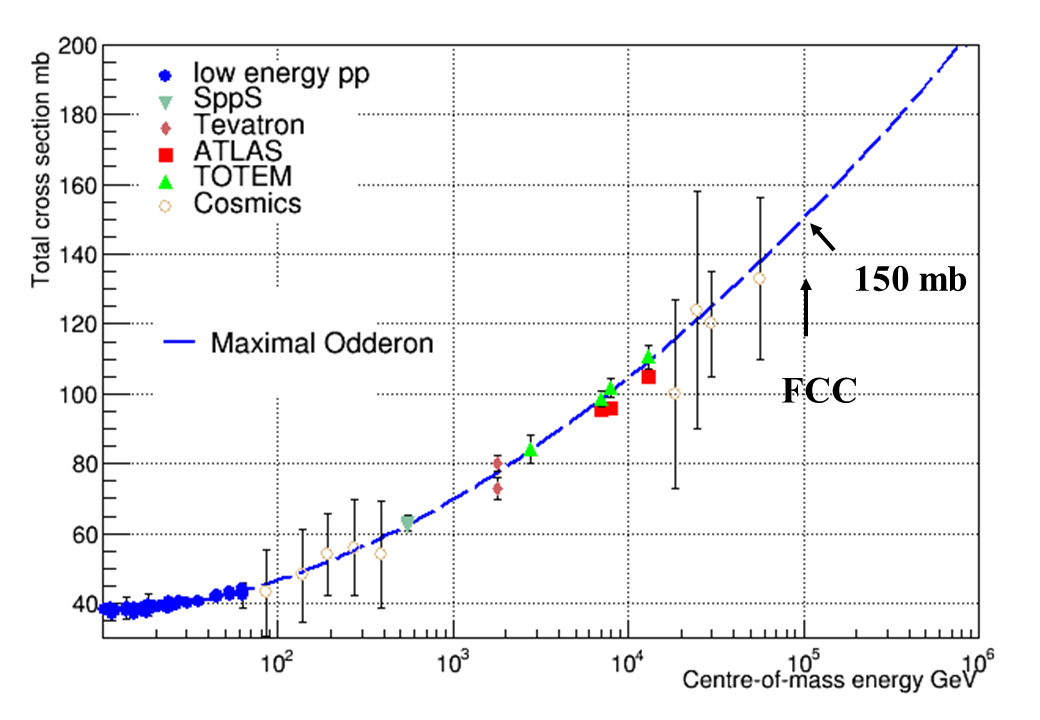} 
\caption{$\sigma_{tot}$}
\label{fp7}
\end{subfigure}
\begin{subfigure}{0.49\textwidth}
\includegraphics[width=0.9\linewidth, height=6cm]{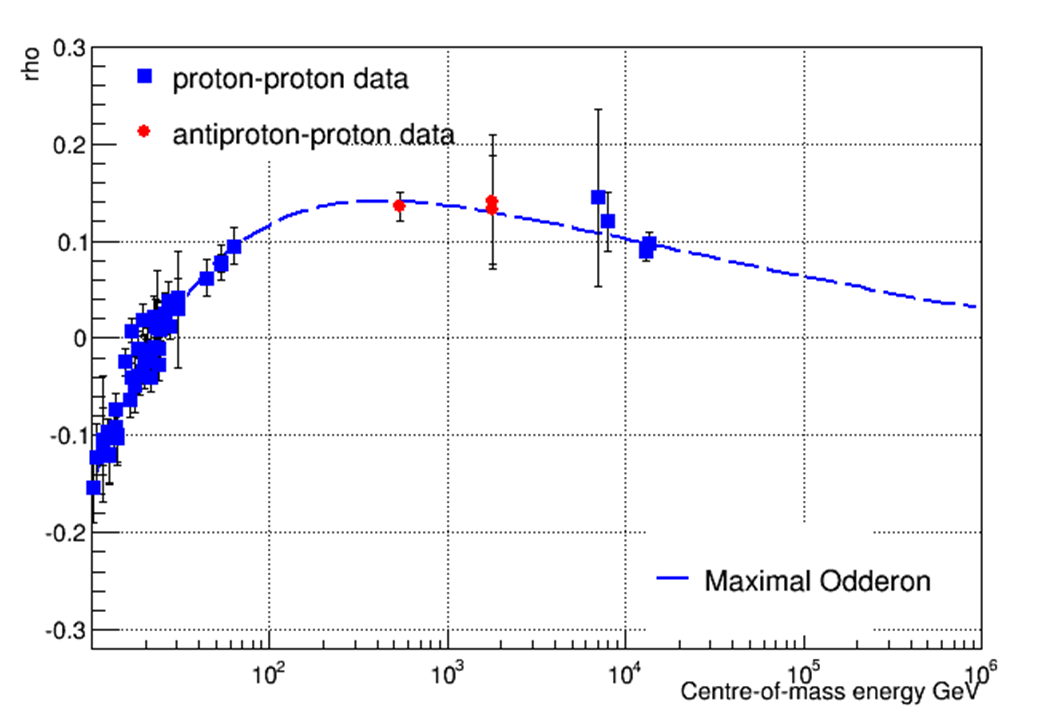}
\caption{$\rho$ }
\label{fp8}
\end{subfigure}
\caption{$\sigma_{tot}$ and $\rho$ as a function of the centre-of-mass energy compared to calculations using the so called Maximal Odderon.  More details are given in the text.}
\label{fig:fmo}
\end{figure}
 For the purpose of this article we will consider the Maximal Odderon in order to see the consequences for  $\sigma_{tot}$ at the FCC.  We have been taking the parameterisation  of FMO from  Ref.~\cite{Martynov:2017zjz} where the model was tuned to the TOTEM data,  whereas ATLAS cross section data at 7 TeV and 8 TeV were discarded from the model tuning.
The result is shown in Fig.\ref{fp7} and Fig.\ref{fp8}. Not surprisingly the Maximal Odderon model end up around 150 mb at the FCC like the other   $ln^{2}(s)$ parameterisations.

Also the authors of Ref.~\cite{PhysRevD.98.074012}   assume an Odderon i.e. both a crossing even and a crossing odd amplitude. They apply a Regge pole approach to the scattering amplitude. Their approach shows quite some resemblance with  the Maximal Odderon approach. The   $\sigma_{tot}$   at the FCC is 150 mb, thus very similar to the FMO model. An interesting feature  of their approach is that the  inelastic profile  in the impact parameter space  $ b$ has a small dip at $b=0$  and in this context one talks about "hollowness"of the proton.

\begin{figure}
\centering
\includegraphics[width=0.9\textwidth]{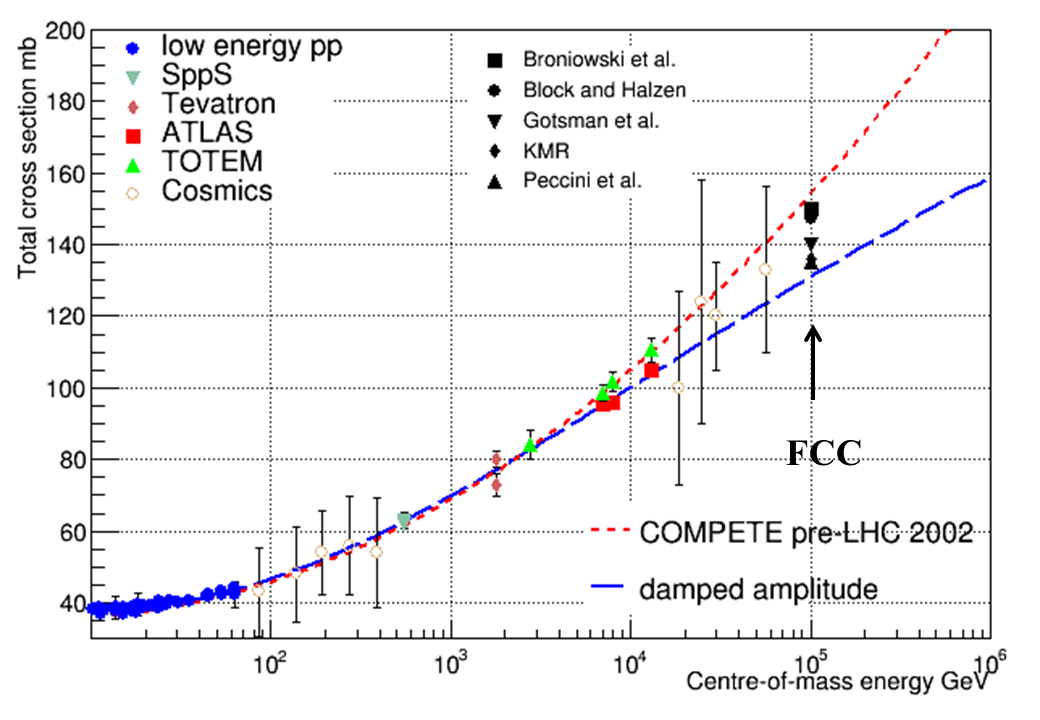}
\caption{\label{fp9}$\sigma_{tot}$ as a function of centre-of-mass energy. The projected limits at the FCC are indicated. Result of different models are also indicated. The corresponding references are found in the text.}
\end{figure}

\section{Conclusion}
We have tried to estimate an upper and lower limit  of  $\sigma_{tot}$ at the FCC. As an upper limit we have taken the "standard"  $ln^{2}(s)$   evolution  that has worked quite  well already since ISR times. To estimate a lower limit we have used dispersion relations assuming that there is no Odderon contribution and that a slower growth of  $\sigma_{tot}$ can be parametrised  as  \(\frac{ln^{2}(s)}{1+\alpha ln^{2}(s)}\) where \(\alpha\) represents a damping factor. We have also looked at several models  with more physics content that describe  quite well available data for $\sigma_{tot}$ and $\rho$ without using an Odderon. In addition we  have considered a couple of models assuming a scattering amplitude which is odd under crossing asymptotically, in other words assuming an Odderon contribution. Here the values found are close to the maximum  "standard" $ln^{2}(s)$   evolution.

The overall situation is summarised in Fig.~\ref{fp9} where are indicated the limits  and the values from the different models mentioned here. It is clear that we can not be exhaustive in our list. Also we only refer to models which in an explicit way give the value of  $\sigma_{tot}$ at 100 TeV, either in a table or in a figure.

From Fig.~\ref{fp9} and from what we have discussed in this note we estimate   that  $\sigma_{tot}$  at FCC will lie inside  the range 130-155 mb. We think it is difficult to be more precise given the general  lack of understanding of non-perturbative QCD. The limits given here can be taken as guide lines for any design considerations at the FCC.

\(\) 
 
%\{\bf\Large Acknowledgements}\\

\bibliographystyle{unsrt}
\bibliography{FCC.bib}
%\printbibliography %hier Bibliographie ausgeben lassen

\end{document}